\documentstyle[mncite,psfig]{mn}

\newcommand{\kms}{km\hspace*{0.35ex}s$^{-1}$}

\newcommand{\oiii}{{[OIII]$\lambda$5007 }}
\newcommand{\oii}{{[OII]$\lambda\lambda$3727 }}
\newcommand{\arcdot}{{$^{\prime\prime}$\hspace*{-0.6ex}.}}

\newcommand{\lesssim}{\raisebox{-0.6ex}{$\,\stackrel
{\raisebox{-.2ex}{$\textstyle <$}}{\sim}\,$}}
\newcommand{\gtrsim}{\raisebox{-0.6ex}{$\,\stackrel
{\raisebox{-.2ex}{$\textstyle >$}}{\sim}\,$}}

\title[3C\hspace{0.35ex}265: AGN-photoionization and jet-cloud
interactions] {Shocks, illumination cones and intrinsic gas structures in
the extreme radio galaxy 3C\hspace{0.45ex}265}

\author[C.\ Sol\'{o}rzano-I\~{n}arrea et\ al.]
{C. Sol\'{o}rzano-I\~{n}arrea$^{1,2}$\thanks{E-mail:
csi@roe.ac.uk}\thanks{Present address: Institute for Astronomy, University
of Edinburgh, Royal Observatory, Edinburgh, EH9 3HJ, UK.}, C. N. Tadhunter$^1$ and
J. Bland-Hawthorn$^3$\\ $^{1}$Department of Physics and Astronomy,
University of Sheffield, Sheffield,   S3 7RH, UK\\ $^{2}$Department of
Physics and Astronomy, University of Leeds, Leeds, LS2 9JT, UK\\
$^{3}$Anglo-Australian Observatory, PO Box 296, Epping, NSW 2121,
Australia}

\begin{document}
\maketitle

\begin{abstract}

We present deep, narrow-band and continuum images of the powerful
high-redshift radio galaxy 3C\hspace*{0.35ex}265 (z=0.811), taken with the
TAURUS Tunable Filter on the William Herschel Telescope, together with
detailed long-slit spectroscopic observations along  the axis defined by
the UV/optical emission elongation. The deep images reveal the existence
of cones in the ionization structure of 3C\hspace*{0.35ex}265 within
$\sim$7 arcsec (58 kpc) of the nucleus, where the emission-line structure
is not  observed to be closely aligned with the radio axis. This indicates
that anisotropic illumination from the central active nucleus dominates on
a small scale. In contrast, at larger distances ($\gtrsim$ 10 arcsec; 80
kpc) from the nucleus, low-ionization  emission gas is closely aligned
with the radio axis, suggesting that jet-cloud  interactions may become
the dominant mechanism in the line-emitting gas on a larger
scale. Moreover, the presence of a high-velocity cloud at 2.5 arcsec from
the nucleus,  close to the radio axis, indicates that even close to the
nucleus ($\sim$20 kpc)  jet-induced shocks have an important kinematic
effect. However, spectroscopic  analysis of this region reveals that the
ionization state of the high-velocity gas is  similar to or higher than
that of the surroundings, which is opposite to what we  would expect for a
cloud that has been compressed and accelerated by jet-induced  shocks.

Our images show that, while on a large scale the
low-ionization emission-line structures are aligned with the radio
axis, on a smaller scale, where AGN-photoionization dominates, the
highest surface-brightness structure is aligned  with the closest
companion galaxy (misaligned with the radio axis). This suggest  that
much of the emission-line structure reflects the intrinsic gas
distribution,  rather than the ionization pattern imprinted by the
radio jets or by illumination from  the central AGN.

Overall, our results underline the need for a variety of mechanisms to
explain the  properties of the extended emission-line gas in the
haloes of radio galaxies.
   
\end{abstract}

\begin{keywords} 
galaxies: active --- galaxies: individual: 3C\hspace{0.35ex}265 ---
galaxies:  structure --- galaxies: jets --- galaxies: kinematics and
dynamics.
\end{keywords}

\section{Introduction}

One of the most important issues concerning radio galaxies is the
dominant  ionization mechanism of the extended line-emitting gas in
these sources. These emission-line regions extend large distances from
the nucleus, and provide important information about both the origin
of the activity and the origin of the extended gas.  However, the main
physical processes affecting the properties of the extended
emission-line regions (EELR) are not fully understood.

The two most accepted models are: a) photoionization by the central
active galactic  nucleus (AGN), and b) shock-ionization by the
interactions between the radio- and  line-emitting structures, the
so-called `jet-cloud interaction' model.  (See \pcite{cliverew2001}
for a review.)

While most (but not all) low-redshift radio galaxies appear to be
consistent with  AGN-photoionization, as we move to higher redshifts
there is increasing evidence  for shocks. Many distant radio galaxies
present highly disturbed kinematics, mostly  along the radio axis
\cite{mccarthy96}, and highly collimated optical/UV structures
aligned along the radio axis \cite{mccarthy87,best96}.

In addition to the general alignments, detailed study of a sample of
z$\sim$1  radio galaxies has revealed a strong evolution of the
emission-line properties  with radio size \cite{best2000}: the
emission-line regions of small radio sources  show disturbed
kinematics and emission-line ratios in agreement with shock
predictions; while larger radio sources appear more quiescent and
present  emission-line ratios consistent with
AGN-photoionization. However, recent kinematic  studies of a sample of
z$\lesssim$0.8 radio galaxies have provided evidence that  shocks have
an important effect on the emission-line properties in all sources,
even  those in which the radio structures are on a much larger scale
than the  emission-line structures \cite{carmen2001}.

It seems clear that a combination of both AGN-photoionization and
shock-ionization  is required to explain the observed properties in the
emission-line regions of radio  galaxies over the whole range of redshifts
and radio power. However, the balance  between these mechanisms is not yet
clear. In order to address this issue, we require  complete maps of both
the kinematics and ionization structure in the extended  haloes of radio
galaxies, and not only spectroscopic studies along a preferred  direction
(e.g. radio axis), which offer a limited view of the properties of the
emission-line gas in these sources. To this end, in this paper we present
deep  emission-line imaging of the powerful high-redshift radio galaxy
3C\hspace*{0.35ex}265, taken with  the TAURUS Tunable Filter (TTF), which
are aimed at determining the dominant physical  mechanisms. While the
published ground-based images of high-redshift radio galaxies are
relatively shallow, and the higher-resolution Hubble Space Telescope (HST)
images are insensitive to low-surface brightness structures, the TTF is
sensitive enough to detect faint  structures in the galaxy halo. In
addition, the TTF also allows a narrow bandpass, which,  for our observations,
was tuned to observe the high-velocity gas \cite{tadhunter91},  as well as
the unshifted component, therefore permitting a two-dimensional kinematic
analysis of the galaxy.

To supplement the TTF images, we also present long-slit spectroscopic
observations  of 3C\hspace*{0.35ex}265 taken along
PA\hspace*{0.35ex}145$^{\circ}$, the axis  defined by the UV/optical
emission elongation. Preliminary results of these  observations were
presented in \scite{tadhunter91}. Here we present a more detailed
analysis.

Throughout this paper a Hubble constant of H$_{0}$ = 50 km s$^{-1}$
Mpc$^{-1}$ and a density parameter of $\Omega_{0}$~=~1 are assumed,
resulting in an angular scale of  8.25 kpc~arcsec$^{-1}$ for
3C\hspace*{0.35ex}265.

\section{Previous observations of 3C\hspace*{0.35ex}265}

3C\hspace*{0.35ex}265 is a large (78 arcsec; 643 kpc) radio source,
with a redshift  z=0.811 (see radio map in \pcite{fernini93}). Its
extreme emission-line luminosity --- an order of magnitude brighter
than other radio galaxies at the same redshift
\cite{thesismccarthy88} --- makes 3C\hspace*{0.35ex}265 unique for the
study of  emission-line properties in distant radio galaxies.

Unlike other powerful high-z radio galaxies
(e.g. 3C\hspace*{0.35ex}368),  3C\hspace*{0.35ex}265 is not a good
example of the alignment effect. The rest-frame  UV emission is
misaligned by approximately 35$^{\circ}$ relative to the radio axis.
Nevertheless, this misalignment is still consistent with the large
half-opening  angles ($\sim$~45$^{\circ}$ -- 60$^{\circ}$) for the
ionization cones predicted by  the unified schemes for radio sources
(e.g. \pcite{barthel89}).

3C\hspace*{0.35ex}265 presents strongly polarized rest-frame UV
continuum and  broad MgII\hspace*{0.5ex}$\lambda\lambda$2799 emission,
with the polarization  position angle oriented perpendicular to the
extended UV emission rather than  to the radio axis
\cite{jannuzi91,dey96,di-serego-ali96}. This provides strong  evidence
for the extended gas being illuminated by a powerful hidden quasar in
the nucleus of 3C\hspace*{0.35ex}265. In addition, a more recent
analysis of the  ionization state of the emission-line gas in
3C\hspace*{0.35ex}265 shows that  the near-UV emission-line ratios are
consistent with AGN-photoionization  (Best et al. 2000). Therefore, at
first sight, 3C\hspace*{0.35ex}265 appears to be  one of the best high
redshift examples of an object dominated by
AGN-photoionization.\nocite{best2000}

However, \scite{tadhunter91} reports the discovery of high-velocity
gas in the  extended emission-line region of
3C\hspace*{0.35ex}265. The presence of this high-velocity gas suggests
that jet-induced shocks may also contribute to the  ionization of the
ambient gas at some level.

\section{Observations, data reduction and analysis}

\subsection{Optical spectroscopy}

Long-slit spectroscopic observations were carried out on the night
23/01/91,  using the red arm of the ISIS double spectrograph on the
4.2m William Herschel  Telescope (WHT) on La Palma (Spain). The EEV2
CCD detector was used, giving a pixel  scale of 0.327 arcsec/pix. The
2.16 arcsec slit was oriented along the axis  defined by the
elongation of the optical/UV emission
(PA\hspace*{0.35ex}145$^{\circ}$), roughly perpendicular to the
observed B-band  polarization position angle \cite{jannuzi91}, and
misaligned by $\sim35^{\circ}$ relative to the radio axis. Two individual
spectra were coadded to produce a single  spectrum with a exposure
time of 3600 seconds. Details of the observations are  presented in
\scite{tadhunter91}.

\begin{table*}
\begin{center}
\begin{tabular}{lccclc}\hline \\
 &{\bf Blocking Filter} &{\bf Etalon}&{\bf t$_{exp}$} & {\bf Surface Brightness Limit} & {\bf Seeing (FWHM)} \\
 &$\lambda_{c}$(\AA)/$\Delta\lambda$(\AA) & $\lambda_{c}$(\AA)/$\Delta\lambda$(\AA)& (s) & (erg cm$^{-2}$ s$^{-1}$
 arcsec$^{-2}$) & (arcsec)\\ \\ \hline \\
     Continuum       & 8122/330 &   ---   & 2$\times$900 & \hspace*{0.8cm}$\sim 3 \times 10^{-17}$ $^{(}$*$^{)}$ & 1.17$\pm$0.10\\ 
 $[$OII$]$ low-vel. & 6682/210 & 6742/20 & 2$\times$900 & \hspace*{0.8cm}$\sim 1 \times 10^{-17}$ & 1.16$\pm$0.05\\
 $[$OII$]$ high-vel. & 6682/210 & 6769/20 & 900  &\hspace*{0.8cm}$\sim 1.6 \times 10^{-17}$ & 1.20$\pm$0.06\\ 
 $[$OIII$]$ low-vel. & 9094/400 & 9052/36 & 2$\times$900 &\hspace*{0.8cm}$\sim 1.6 \times 10^{-17}$ & 1.18$\pm$0.06\\ 
 $[$OIII$]$ high-vel.& 9094/400 & 9090/36 & 2$\times$900 &\hspace*{0.8cm}$\sim 3 \times 10^{-17}$ & 1.07$\pm$0.09\\ \\ \hline
\end{tabular}
\caption{Log of the TTF imaging observations for 3C\hspace{0.35ex}265. The
second and third columns give the central wavelength/bandwidths for the blocking
filters and the etalon, respectively. \ $^{(}$*$^{)}$Total surface brightness integrated across the 
bandwidth of the broad-band filter (330\AA).}
\end{center}
\label{ttflog}
\end{table*}

The reduction of the long-slit spectra was performed using the IRAF
software  package, following the usual steps: bias-subtraction,
cosmic-ray removal,  flat-fielding, wavelength calibration,
atmospheric extinction correction, flux  calibration and sky
subtraction. A relative flux calibration error of $\sim$9~\%  across
the wavelength range was found by comparing different
spectrophotometric  standard star observations. After the basic
reduction procedure, the  two-dimensional spectra were corrected for
any distortion along the spectral  direction, using the Starlink
FIGARO package. Finally, the spectra were shifted  to the rest frame
of the galaxy before the analysis. Then, the extracted
one-dimensional spectra were analysed using the Starlink DIPSO
spectral analysis  package. Gaussians were fitted to the emission-line
profiles. The measured  linewidths were corrected for the spectral
resolution of the instrument, which  was derived using the night-sky
and arc lines, and found to be 15.60$\pm$0.16~\AA.  Since the airmass
during the observations of 3C\hspace*{0.35ex}265 was  1.03, the
effects of differential atmospheric refraction can be neglected.  The
data have not been corrected for Galactic reddening, given that
3C\hspace*{0.35ex}265 does not have a low Galactic latitude, and
therefore the extinction due to dust in our Galaxy is not large
[E(B-V)=0.023 mag; taken from NASA/IPAC Extragalactic Database (NED)].

\subsection{Optical imaging}

Emission-line and continuum observations of 3C\hspace*{0.35ex}265 were
taken on the  night 22/01/98 using the TTF on the WHT on La Palma
(Spain). A full description of the  TTF is given in Bland-Hawthorn \&
Jones (1998a,b)\nocite{bland-haw98,bland-haw98spie}.  Use of the f/2
camera of TAURUS and the Tek5 CCD detector resulted in a pixel scale
of 0.557 arcsec/pix. A log of the observations is presented in
Table~1.  The etalon was tuned to the redshifted wavelengths of the
low- and high-velocity  components of both \oii and \oiii emission
lines (see \pcite{tadhunter91} and  Section~\ref{res:kinemat} for
details of the shifted components). Observations of the
spectrophotometric standard star SP1337+705 and dome flats were taken
at each  wavelength setting. Estimates of the seeing FWHM were derived
by fitting the spatial  profiles of several stars in the field. No
significant variations in the seeing were  found between the different
filter observations (see Table~1 for details).

The reduction of the TTF images was performed using IRAF and following
the usual  steps. After subtracting the bias (using the overscan
region), the images were  divided by the corresponding normalized
flat-field frame, and cosmic rays were  manually removed using the
task {\footnotesize IMEDIT}. The different filter images  were
registered to a reference image (chosen to be that with the lowest
airmass) to  an accuracy of better than 0.1 pixels, using several
stars in the frames (task  {\footnotesize IMALIGN}). Those images
taken with the same wavelength setting were  coadded together. The
resulting frames were corrected for atmospheric extinction,
flux-calibrated and sky-subtracted. Because only one
spectrophotometric standard  star was observed during the
3C\hspace*{0.35ex}265 observations, absolute flux  calibration errors
could not be estimated directly from the TTF images. To estimate  the
uncertainty in the flux calibration, we compared the reduced TTF
images with the  long-slit spectra along
PA\hspace*{0.5ex}145$^{\circ}$, by simulating the position  and width
of the slit on the TTF frames. It was found that the [OII]/[OIII] line
ratio obtained from the TTF images agreed with that derived from the
long-slit  spectra to within $\pm$15~\% for radii less than 7 arcsec
from the nucleus,  increasing to $\pm$40~\% at distances of 7 to 10
arcsec from the nucleus. The  surface brightness of the faintest
structures visible in the images is given in  Table~1.

\section{Results}

\subsection{Optical spectroscopy}

\subsubsection{Emission-line kinematics}
\label{res:spec-kin}

Kinematic studies of the line-emitting gas in 3C\hspace*{0.35ex}265,
based on  long-slit spectroscopic observations (e.g. \pcite{dey96}),
show an ordered velocity  profile with small velocity shifts
($\Delta$$v\lesssim$300 \kms) relative to the nucleus, at least for
the strongest emission-line components. This is consistent with
gravitational  motions in the potential of a single galaxy
\cite{tadhunter89b,baum90}.

\begin{figure*}
\centerline{ \psfig{figure=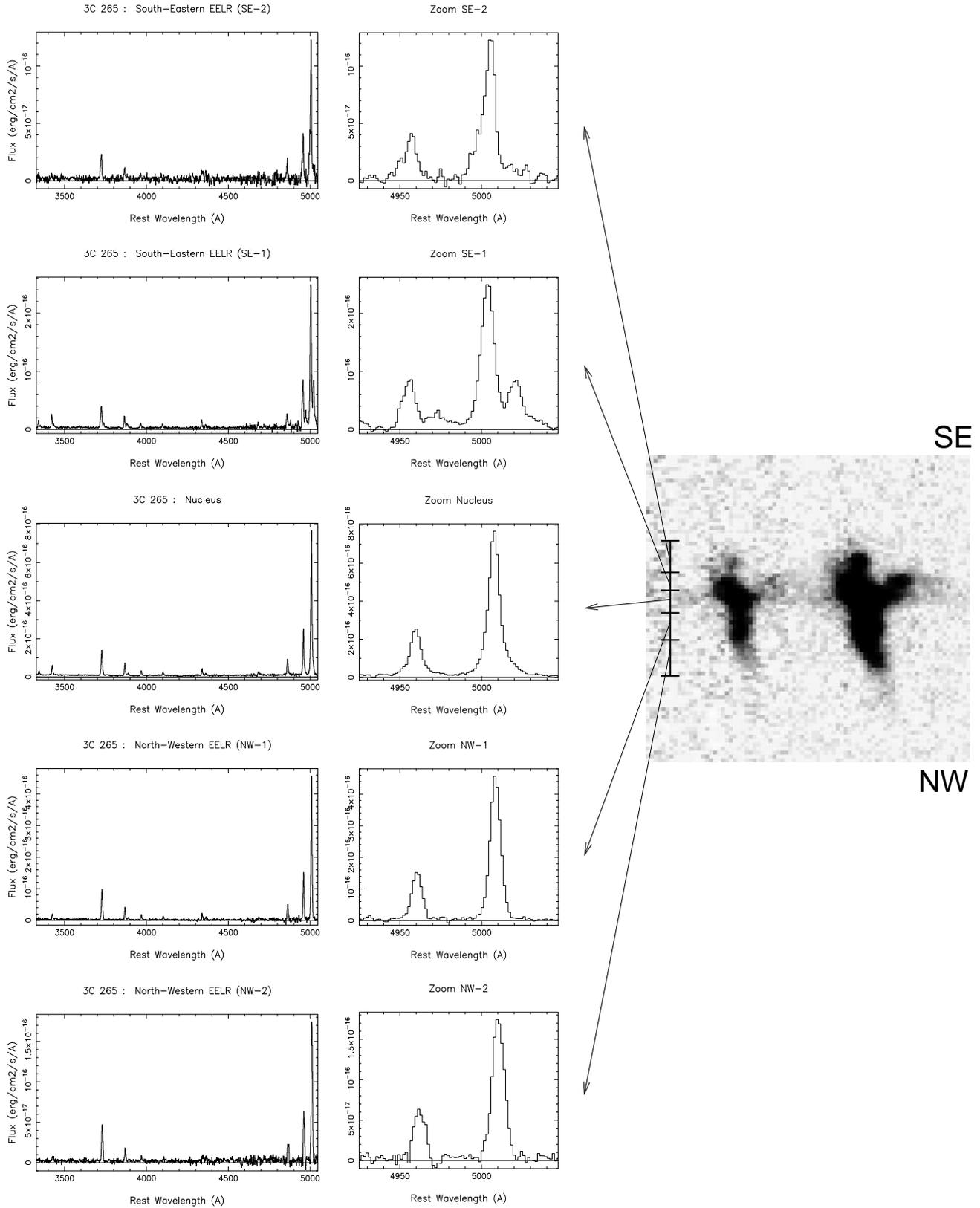,angle=-90,width=17.5cm}}
\caption[]{Integrated spectra of the different regions in
3C\hspace*{0.35ex}265, together with a zoom of the [OIII] doublet, and
2-D spectrum showing the doublet  [OIII]$\lambda\lambda$4959,5007
(4925 -- 5047 \AA, 25.5 arcsec along the
PA\hspace*{0.5ex}145$^{\circ}$ slit).}
\label{specfig}
\end{figure*}

\begin{table*}
\footnotesize
\begin{center}
\begin{tabular}{lcccccc}\hline \\
\vspace*{0.7ex}
 &{\bf SE-2 EELR}&\multicolumn{2}{c}{\bf SE-1 EELR}&{\bf Nucleus}&{\bf NW-1 EELR}&{\bf NW-2 EELR}\\ 
\vspace*{1ex}
 & (3\arcdot8 SE) & \multicolumn{2}{c}{(1\arcdot5 SE)} &(0\arcdot3 NW) & (2\arcdot4 NW) & (5\arcdot0 NW) \\
{\hspace{0.5ex}\bf Line}& &{\bf low-vel.}&{\bf high-vel.}& & & \\ \\ \hline \\

 $[$NeV$]\lambda3346$ \dotfill &           ---      & 0.27$\pm$0.04 & 0.40$\pm$0.15 & 0.24$\pm$0.02 & 0.11$\pm$0.02 & --- \\
 $[$NeV$]\lambda3426$ \dotfill &  $<$\hspace{0.3ex}0.37 & 0.75$\pm$0.11 & 1.11$\pm$0.43 & 0.67$\pm$0.04 &
 0.31$\pm$0.03 & $<$\hspace{0.3ex}0.21 \\
 $[$OII$]\lambda\lambda$3727 \dotfill & 2.20$\pm$0.46 & 1.85$\pm$0.19 & 1.18$\pm$0.36 & 2.01$\pm$0.10 & 2.24$\pm$0.17 & 2.16$\pm$0.30 \\
 $[$NeIII$]\lambda$3869      \dotfill& 0.94$\pm$0.21 & 0.86$\pm$0.10 & 0.99$\pm$0.38 & 0.77$\pm$0.04 & 0.77$\pm$0.06 & 0.64$\pm$0.09 \\
 H8+HeI\hspace{0.4ex}3886 \dotfill&    0.27$\pm$0.10 &     ---       &      ---      & 0.21$\pm$0.03 & 0.15$\pm$0.03 & 0.12$\pm$0.04 \\
 $[$NeIII$]\lambda$3967   \dotfill&    0.30$\pm$0.07 & 0.27$\pm$0.03 & 0.31$\pm$0.12 & 0.24$\pm$0.02 & 0.24$\pm$0.02 & 0.20$\pm$0.03 \\
 H$\delta$ \dotfill&       ---      & 0.23$\pm$0.05 &      ---      & 0.24$\pm$0.02 & 0.26$\pm$0.03 & 0.16$\pm$0.05 \\
 H$\gamma$ \dotfill&       ---      & 0.49$\pm$0.07 &      ---      & 0.45$\pm$0.03 & 0.37$\pm$0.04 & 0.37$\pm$0.09 \\
 $[$OIII$]\lambda$4363 \dotfill&          ---      & 0.24$\pm$0.05 &      ---      & 0.20$\pm$0.02 & 0.17$\pm$0.03 & 0.25$\pm$0.07 \\
 HeII\hspace{0.4ex}$\lambda$4686 \dotfill&            ---      & 0.32$\pm$0.07 &      ---      & 0.27$\pm$0.04 & 0.17$\pm$0.05 & --- \\
 H$\beta$ \dotfill&   1.00$\pm$0.21 & 1.00$\pm$0.11 & 1.00$\pm$0.29 & 1.00$\pm$0.05 & 1.00$\pm$0.08 & 1.00$\pm$0.14 \\
 $[$OIII$]\lambda$4959 \dotfill&     3.85$\pm$0.88 & 3.62$\pm$0.40 & 2.74$\pm$0.86 & 3.54$\pm$0.23 & 3.16$\pm$0.25 & 2.44$\pm$0.36 \\
 $[$OIII$]\lambda$5007 \dotfill&     12.57$\pm$2.66& 10.86$\pm$1.19 & 8.21$\pm$2.58 & 11.51$\pm$0.61 & 9.88$\pm$0.76 & 7.88$\pm$1.10 \\ \\ \hline \\ 
H$\beta$ flux   & & & & & \\
 (10$^{-16}$\hspace{0.3ex}erg\hspace{0.4ex}cm$^{-2}$\hspace{0.3ex}s$^{-1}$)&  1.76$\pm$0.36 & 3.71$\pm$0.39 & 1.12$\pm$0.33
 & 11.50$\pm$0.58 & 6.69$\pm$0.51 & 3.73$\pm$0.52 \\ \\ \hline \\
$[$OIII$]\lambda$5007 Velocity & & & & & \\
  (\kms) & --90$\pm$8 $^{(}$*$^{)}$ & --222$\pm$3 & +800$\pm$11 & +10$\pm$2 $^{(}$*$^{)}$ & +61$\pm$2 & +176$\pm$5 \\ \\ \hline \\
\end{tabular}
\end{center}
\caption{Emission-line fluxes of 3C\hspace*{0.35ex}265, including the south-eastern EELR `SE-2' (2.94$\times$2.16 arcsec$^{2}$ aperture centred at 3.8 arcsec SE of
the continuum centroid), the south-eastern EELR `SE-1' (1.63$\times$2.16 arcsec$^{2}$ aperture centred at 1.5 arcsec SE of the continuum centroid),
the nucleus (1.96$\times$2.16 arcsec$^{2}$ aperture centred at 0.3 arcsec NW of the continuum centroid), the north-western EELR `NW-1' (2.29$\times$2.16
arcsec$^{2}$ aperture centred at 2.4 arcsec NW of the continuum centroid) and the north-western EELR `NW-2' (2.94$\times$2.16 arcsec$^{2}$ aperture
centred at 5.0 arcsec NW of the continuum centroid). The velocity of the [OIII]$\lambda$5007 emission-line for each aperture with respect to the velocity 
at the continuum centroid is also given. $^{(}$*$^{)}$These velocities correspond to the main, narrow component of the fit.}
\label{tabflux265}
\end{table*}

\begin{figure}
\centerline{ \psfig{figure=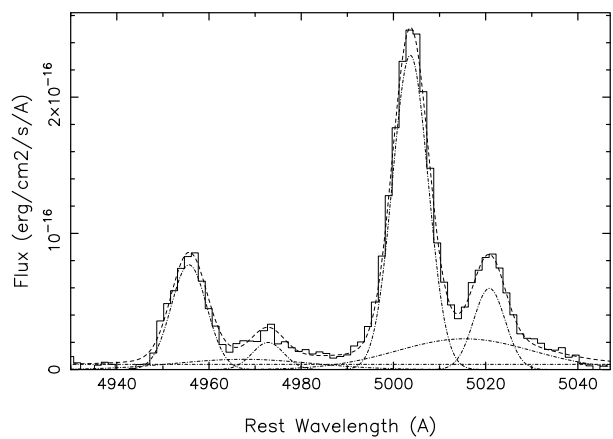,width=9cm,angle=-90}}
\caption[]{[OIII]$\lambda\lambda$4959,5007 emission-line profiles
(solid line) from  the long-slit spectrum: 1.63$\times$2.16
arcsec$^{2}$ aperture centred at 1.5 arcsec  SE of the continuum
centroid. The multicomponent Gaussian fits (dot-dashed-dot lines)  and
the total fit (dashed line) are also plotted.}
\label{o3fit}
\end{figure}

Our reanalysis of the data presented in \scite{tadhunter91} shows that the
emission-line profiles can be fitted by single Gaussians in the
north-western EELR,  but in the nucleus and south-eastern EELR the
profiles are more complex (see  Fig.~\ref{specfig}). The observed broad
wings to the [OIII]$\lambda\lambda$4959,5007  profiles in these regions
show the existence of an underlying broad component in addition to the
narrow component present in the nucleus and the two SE regions, as well as
a high-velocity narrow component which is strongest in the SE-1 EELR ---
region studied by \scite{tadhunter91}. In Fig.~\ref{o3fit} we show the
[OIII] emission-line profiles (solid line) of the SE-1 region where the
split components are observed, corresponding to a 1.63$\times$2.16
arcsec$^{2}$ aperture centred at  1.5 arcsec south-east of the continuum
centroid. Initially, a two-component Gaussian  fit for each line of the
doublet was considered, but the emission-line  profiles could not be
reproduced by this fit. Thus, a three-component Gaussian fit  was
considered (dot-dashed-dot) to fit the red broad wing seen in the [OIII]
line, suggestive of the existence of an underlying broad component. The
total fit  is indicated by a dashed line. At the location of the
splitting, this best fit consists of a  blue-shifted low-velocity component at
--222$\pm$3 \kms, a red-shifted high-velocity component at  +800$\pm$11
\kms, and a very broad red-shifted underlying component with a velocity
of +500$\pm$100 \kms \ and a linewidth of 2000$\pm$180 \kms (FWHM). This
provides a better fit than the 
three narrow components suggested by \scite{tadhunter91}. The
velocity shifts of  the three components are given with respect to the
velocity at the continuum  centroid. The low- and high-velocity
components have linewidths close to the  spectral resolution limit.

\subsubsection{Emission-line spectra}
\label{res:spec-fluxes}

Integrated spectra of five spatial regions along the slit for
3C\hspace*{0.35ex}265,  together with a two-dimensional spectrum
showing the [OIII]$\lambda\lambda$4959,5007  doublet, are presented in
Fig.~\ref{specfig}. The integrated fluxes of the observed
emission-lines in these regions, normalized to the corresponding
observed-frame  H$\beta$ flux, are given in
Table~\ref{tabflux265}. The line ratios have not been  corrected for
intrinsic reddening, given that the intensities of H$\delta$ and
H$\gamma$ relative to H$\beta$ are not significantly lower than
expected for Case B  recombination \cite{osterbrock89}, suggesting that
the intrinsic reddening is  not large.

\subsection{Optical imaging}

\subsubsection{Emission-line and continuum structures}

\begin{figure*}
\centerline{ \psfig{figure=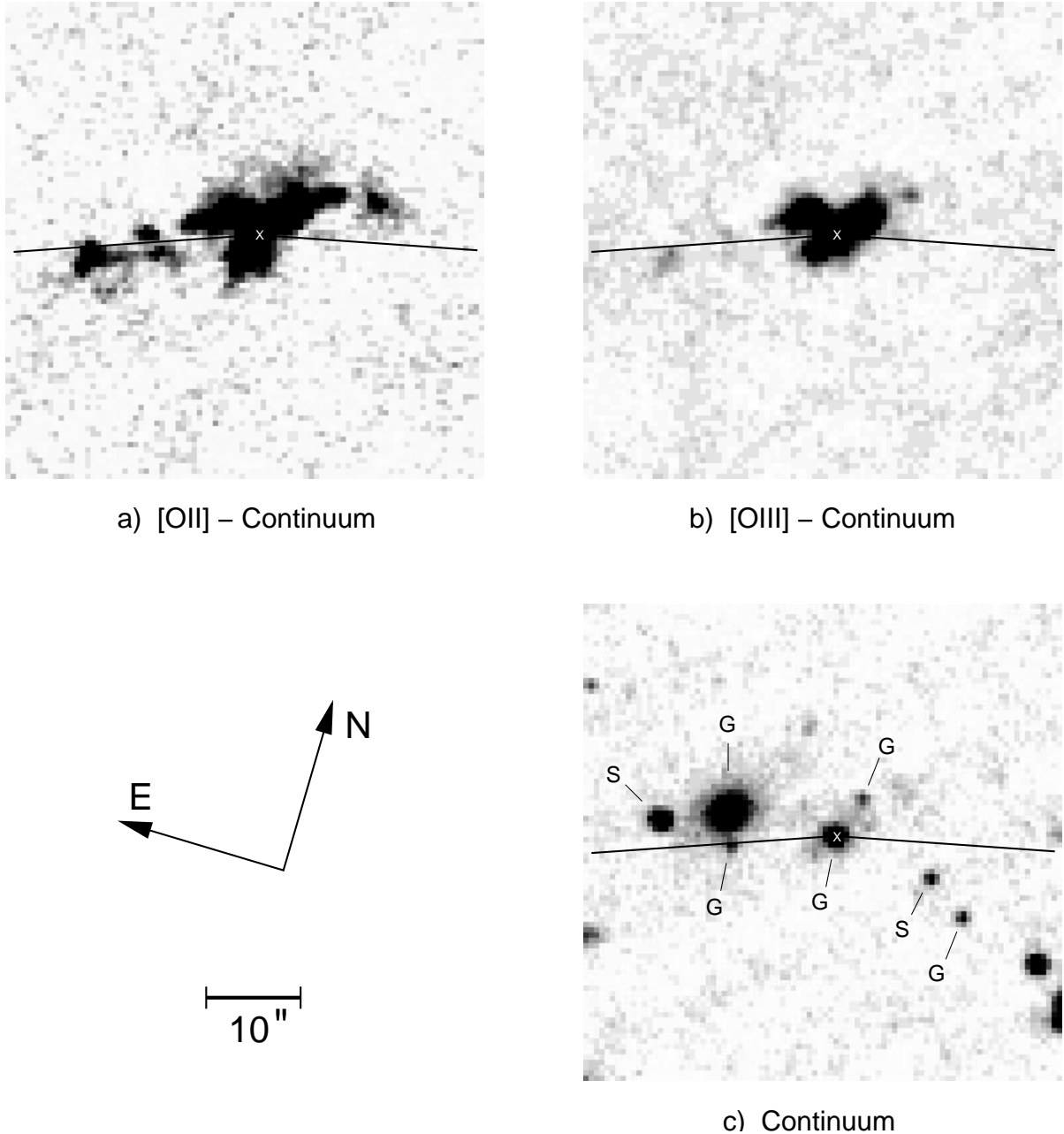,angle=-90,width=16cm}}
\caption{TTF images of 3C\hspace*{0.35ex}265: a) \oii emission-line
image for the low-velocity component; b) \oiii emission-line image for
the low-velocity component;  c) Continuum image centred on
(rest-frame) 4485 \AA. On the basis of examination of high resolution
HST images, the objects detected in the continuum image (c) have been
classified into galaxies (G) and stars (S). All three images cover
50.13 arcsec  (414 kpc) in both directions. The galaxy continuum
centroid is indicated by a  white `x', and the solid line represents
the radio axis.}
\label{o2o3cont}
\end{figure*}

Fig.~\ref{o2o3cont} presents deep \oii, \oiii and continuum images of
3C\hspace*{0.35ex}265, showing the spectacular emission structure of
this  object. The galaxy continuum centroid is indicated by a white
`x', and the solid  lines represent the axes of the large scale radio
emission on either side of the nucleus \cite{fernini93}.

Image (a) shows the continuum-subtracted [OII] emission-line structure
for the  low-velocity component. Before the continuum subtraction, the
continuum image was  scaled to the [OII] image according to the
corresponding bandwidths of the  broad-band filter and etalon,
respectively. It can be seen that the emission extends  over more than
35 arcsec (290 kpc) across the galaxy, confirming previous results
presented by \scite{rigler92} and \scite{mccarthy95}. On the eastern
side of the  galaxy the [OII] emission extends up to approximately 20
arcsec (165 kpc) from the  centre of the galaxy, and is closely
aligned along the radio axis on a large scale, between 10  and 20
arcsec (82 -- 165 kpc) from the nucleus. However, closer to the
nucleus, no clear alignment can be observed between the emission and
radio structures. In fact,  within a radius of 10 arcsec the highest
surface-brightness [OII] emission appears  to be misaligned by
$\sim$30 -- 40$^{\circ}$ with respect to the radio axis. Note  also
the presence of significant flux perpendicular to the radio axis in
the central  regions (r $\lesssim$ 7 arcsec; 58 kpc).

Of particular interest is the `hole' in the emission structure located
between  10 and 12 arcsec (82 -- 100 kpc) along the radio axis on the
eastern side of the  nucleus. Note the remarkable coincidence of the hole
lying exactly on the radio axis,  suggesting that it is due to the passage
of the radio jet through the cloud.  Faint linear features can also be
noticed along the radio axis at $\sim$13 arcsec  (107 kpc) to the east of
the nucleus, and possibly at $\sim$7 arcsec (58 kpc) to the west of the
nucleus.

The continuum-subtracted [OIII] emission-line structure for the
low-velocity  component is presented in image (b). The continuum
subtraction was performed in the same way as for the [OII] image. The
emission structure in [OIII] is much more  centrally concentrated than
the [OII] structure (this is not due to the relative depth of the
images).  The high-surface-brightness  structures do not show a clear
alignment with the radio axis. Only a faint [OIII] emission region can
be seen along the radio axis, $\sim$18 arcsec (150 kpc) to the  east
of the nucleus. There are indications that the possible linear feature
observed in the [OII] image at $\sim$7 arcsec (58 kpc) to the west of
the nucleus, may be also detected at the same location  along the
radio axis in the [OIII] structure.

It can be seen that the [OII] and [OIII] emission-line structures are
consistent with a large ($\sim60^{\circ}$) half-opening angle for the
ionization bicone predicted  by the unified schemes. However, in both
images, significant flux is also detected  outside the cones,
perpendicular to the radio axis, in the nuclear regions. Note also the
apparent ``hollowing out'' of the cones, which could be the result of
outflows  induced by the central AGN. A similar edge-brightened
biconical structure is seen on a much smaller scale in 
Cygnus\hspace*{0.35ex}A \cite{tadhunter99}.

Image (c) shows the continuum emission of 3C\hspace*{0.35ex}265. The
central  (rest-frame) wavelength of the filter was 4485~\AA, with a
bandwidth of 330~\AA.  Therefore, any important emission-line
contamination was avoided. This image shows  a central compact galaxy,
with a companion galaxy at 5 arcsec (41 kpc) to the  north-north-west,
which is also detected in infrared images \cite{best97} and presents
colours consistent with the same redshift as 3C\hspace*{0.35ex}265
(Best,  private communication). The bright galaxy at 12 arcsec to the
east of  3C\hspace*{0.35ex}265 is a foreground elliptical galaxy at
redshift z=0.392  \cite{smith79}. There is also a faint galaxy at
$\sim$11 arcsec (90 kpc) to the east of the nucleus, coincident with
the emission-line structures aligned along  the radio axis. The
colours of this galaxy do not give an unambiguous estimate of  its
redshift: it could be either at the redshift of 3C\hspace*{0.35ex}265
or at the redshift of the brighter foreground elliptical galaxy (Best,
private communication).
 
\subsubsection{Emission-line kinematics: high-velocity gas}
\label{res:kinemat}

\begin{figure}
\centerline{ \psfig{figure=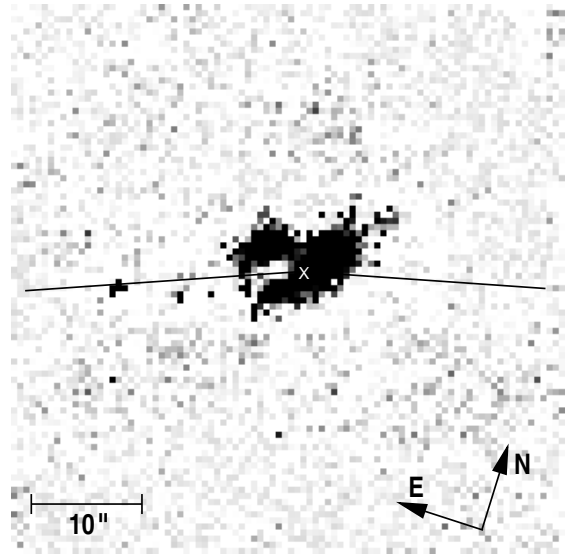,angle=-90,width=7.5cm}}
\caption{Image of 3C\hspace{0.35ex}265 showing the ratio between the
low-    and the high-velocity components to the \oiii emission
line. The high-velocity gas  is represented by white. The continuum
centroid is indicated by a `x', and the solid  line represents the
radio axis. Note the close alignment between the high-velocity  cloud
to the east of the nucleus, and the radio axis.}
\label{lhvel}
\end{figure}

As mentioned above (Section~\ref{res:spec-kin}), we find that the
kinematic structure  of 3C\hspace*{0.35ex}265 to the south-east of the
nucleus along  PA\hspace*{0.5ex}145$^{\circ}$, can be resolved into
three components: two  narrow split components, moving at --222 and
+800 \kms \ relative to the continuum centroid, and a very broad
component with a velocity shift of +500 \kms. Our new TTF  images
provide more information about the location of the +800 \kms \
high-velocity  cloud.

In Fig.~\ref{lhvel} we show the ratio between the low- (--222 \kms) and
the high-  (+800 \kms) velocity filter images for the \oiii emission
line. The high-velocity  gas is represented by white, and black indicates
where the low-velocity component  dominates. The solid line represents the
radio axis and the continuum centroid is indicated by a `x'. Although the
wavelength overlap in the Lorentzian  profiles of the TTF at the two
etalon settings precludes the determination of an accurate flux ratio for
the high- relative to the low-velocity component, this image does allow
the spatial position of the high-velocity component to be determined
accurately.

We find that the high-velocity cloud has a spatial extent of
(1.7$\pm$0.3)\hspace*{0.3ex}$\times$\hspace*{0.3ex}(2.2$\pm$0.3)
arcsec$^{2}$  (14$\times$18 kpc$^{2}$) and is centred at 2.46$\pm$0.13
arcsec (20 kpc) SE of the  continuum centroid of the galaxy, along
PA\hspace*{0.5ex}101.9$^{\circ}\pm$\hspace*{0.4ex}2.0$^{\circ}$. Given
that the  radio axis is along
PA\hspace*{0.5ex}112.0$^{\circ}\pm$\hspace*{0.4ex}0.5$^{\circ}$  on the
eastern side of the radio core (see \pcite{fernini93} for radio map), we
find  that the centroid of the high-velocity cloud is located at a
(perpendicular)  distance of 0.43$\pm$0.09 arcsec (3.5 kpc) north from the
radio axis, subtending  from the galaxy continuum centroid an angular
offset of  10.1$^{\circ}\pm$\hspace*{0.4ex}2.0$^{\circ}$ from the
radio axis on the eastern side of the nucleus. This  close alignment 
strongly suggests that the high-velocity
gas is associated with the  radio jet axis.

\subsubsection{Ionization}
\label{res:ioniz}

\begin{figure*}
\centerline{ \psfig{figure=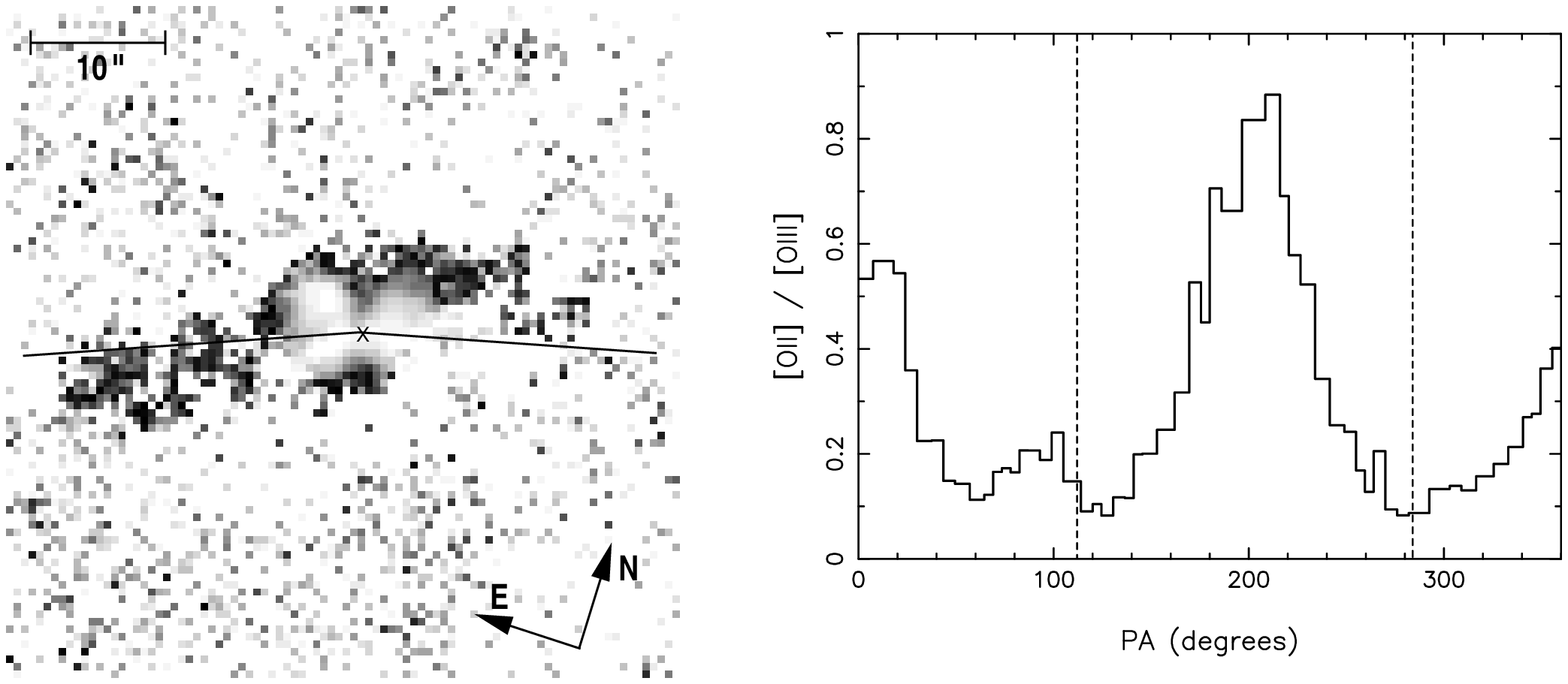,angle=-90,width=17cm}}
\caption{Left: Image of 3C\hspace*{0.35ex}265 showing the \oii/ \oiii
line ratio  for the low-velocity component. White indicates high
ionization, and black indicates  low ionization. The continuum
centroid and the radio axis are indicated by a `x' and  a solid line,
respectively. Right: Azimuthal profile of the line ratio [OII]/[OIII]
in 3C\hspace*{0.35ex}265 (evaluated between 2.23 and 3.34 arcsec
distance from the  nucleus). The vertical dashed lines represent the
radio axes.}
\label{rat+azi}
\end{figure*}

An ionization map of 3C\hspace*{0.35ex}265 is shown in
Fig.~\ref{rat+azi}~(left).  This image is the ratio between the \oii and
the \oiii images, for the low-velocity  component. White indicates high
ionization, and black corresponds to low ionization.  The continuum
centroid and the radio axis are indicated, as in previous figures, by a
`x' and a solid line, respectively.

The ionization state of the nucleus is high, and decreases with radius
from the  centre. At radii smaller than 7 arcsec (58 kpc) from the centre,
the ionization  structure has a `butterfly' shape, which is consistent
with a bicone of 60$^{\circ}\pm$\hspace*{0.4ex}10$^{\circ}$ half-opening
angle, predicted by the  unified schemes. Within the bicone, the
ionization decreases near the edges, and  keeps decreasing in the regions
lying outside it. Note the presence of low-ionization  material located
perpendicular to the radio axis, about 4 arcsec (33 kpc) to the north and
south of the nucleus.

Fig.~\ref{rat+azi}~(right) shows the azimuthal profile of the line
ratio [OII]/[OIII],  extracted from the ionization map and evaluated
between 2.23 and 3.34 arcsec  (18 -- 28 kpc) distance from the
nucleus. The vertical dashed lines represent the  radio axis. This
plot shows how strongly the ionization of the emission-line gas
varies as a function of position angle. A very strong change in the
ionization ratio  is apparent moving away from the radio axis, showing
clear evidence for ionization  cones with half-opening angle of
$\sim$50$^{\circ}$-- 65$^{\circ}$. The ionization  state within the
cones is high (0.1\lesssim[OII]/[OIII]\lesssim0.3), and decreases
sharply towards the edges. It can be seen that gas with lower
ionization state is  present outside the cones around the position
angles 20$^{\circ}$ and 200$^{\circ}$,  with line ratios of
[OII]/[OIII]$\simeq$0.5 and [OII]/[OIII]$\simeq$0.8, respectively.  It
is striking that the low-ionization peaks are separated by
$\sim$180$^{\circ}$, and the radio axes on either side of the nucleus
are roughly mid way between the peaks.
 
On the other hand, as we move to larger distances from the central
region, the  `butterfly' shape in the ionization map vanishes, and
only low-ionization emission  regions can be seen in the
structure. The low-ionization gas extends up to 15 arcsec (124 kpc) to
the north-west and, on the eastern side, is closely aligned along the
radio axis, extending up to a radius of $\sim$24 arcsec (200 kpc).

In addition, Fig.~\ref{rat+azi} resolves the long-standing issue
concerning the optical/UV misalignment with the radio axis in
3C265. \scite{dey96} suggested that this misalignment may be due to a
misalignment between the axis of the quasar illumination cone and the
radio  axis, which would have important implications for the
understanding of AGN.  However, 3C265 clearly possesses a broad
($\sim$60$^{\circ}$ half-opening angle)  ionization cone, aligned with
the radio axis, and therefore the optical/UV misalignment must be due
to the intrinsic matter distribution.

\section{Discussion}

\subsection{The dominant ionization mechanism}

As discussed in the introduction, spectroscopic and polarimetric
studies provide  clear evidence for AGN-illumination in the extended
gas of 3C\hspace*{0.35ex}265  (e.g. \pcite{di-serego-ali96}). This is
also supported by a more recent analysis of  the ionization state of
the extended gas of 3C\hspace*{0.35ex}265, based on  long-slit
spectroscopic observations, which shows that the near-UV emission-line
ratios are consistent with photoionization by the central active
nucleus (Best et al.  2000)\nocite{best2000}.

Our results reinforce the evidence for AGN illumination in the central
regions of  3C\hspace*{0.35ex}265. The \oii and \oiii images,
presented in Fig.~\ref{o2o3cont},  show that the main central
structure is elongated along  PA\hspace*{0.5ex}145$^{\circ}$, and
misaligned with the radio axis by approximately  40$^{\circ}$. The
emission-line structure is consistent with anisotropic illumination
by the quasar which illuminates the ambient gas within the predicted
ionization  cones. In this case, the cones are edge-brightened with a
large  ($\sim$60$^{\circ}\pm$\hspace*{0.4ex}10$^{\circ}$) half-opening
angle. Further  evidence for anisotropic illumination is given by the
ionization map presented in  Fig.~\ref{rat+azi}, which shows the
`butterfly' shape of the ionization structure  in the central region,
suggesting the existence of the predicted cones. The high  ionization
state of the gas within the cones is also clear from this image.

We noted in the previous section the presence of gas outside the
putative ionization  cones, perpendicular to the radio axis on both
sides of the nucleus, and our line  ratio map shows that it has a low
ionization state [see Fig.~\ref{rat+azi} (left)].  The nature of this
emission-line gas cannot be explained in terms of anisotropic
illumination from the central active nucleus, but other possibilities
can be  considered. These include: ionization by some back-scattered
light from within the  cones; photoionization by stars;
shock-ionization as a result of the interactions  between streams of
gas during a merger or accretion event \cite{koekemoer94}; and
shock-ionization due to the lateral expansion of the radio cocoon.

Moving to larger distances from the nucleus ($>$10 arcsec; 80 kpc),
the [OII]  emission-line structure is much more tightly aligned with
the radio axis [see  Fig.~\ref{o2o3cont} (a)], which is not expected
on the basis of illumination by the  broad radiation cones predicted
by the unified schemes \cite{barthel89}. On this scale, the close
alignment and high collimation of the structure is similar to that
observed in objects like 3C\hspace*{0.35ex}266, 3C\hspace*{0.35ex}324
and  3C\hspace*{0.35ex}368, in which shock-ionization is suspected
(Best et al. 2000).  Indeed, such close alignment of low-ionization
gas with the radio axis is in  agreement with the jet-cloud
interaction model. The passage of the radio jet drives  strong shocks
into the clouds, which are compressed, ionized and heated; the density
of the clouds will increase resulting in a lower ionization
parameter. As the  post-shock clouds cool, line radiation will be
emitted, particularly strongly in the  low-ionization lines. Detailed
studies of nearby radio galaxies with jet-cloud  interactions revealed
a large [OII]/[OIII] line ratio in the shocked regions  (Clark et
al. 1997, 1998; \pcite{villarmartin99b}).\nocite{clark97,clark98}

The importance of jet-cloud interactions in the very extended gas of
3C\hspace*{0.35ex}265 is also supported by the detection of a `hole'
in the [OII]  extended structure, lying close to the radio axis, about
11 arcsec (90 kpc) to the east  of the nucleus [see
Fig.~\ref{o2o3cont} (a)], suggesting that this is due to the  passage
of the radio jet through the cloud; and also by the presence
of faint linear features along the radio axis, at 13 arcsec
(107 kpc) east from the nucleus in the  [OII] image, and possibly  
at 7 arcsec (58 kpc) west from the nucleus in both  [OII] and 
[OIII] images [see Fig.~\ref{o2o3cont} (a) and (b)].

Overall, these results indicate that the ambient gas in
3C\hspace*{0.35ex}265 is  mainly ionized by the central active nucleus at
relatively small distances  ($\lesssim$7 arcsec; 58 kpc) from the centre
of the galaxy. However, the situation at  larger radii is less
clear. Because of the r$^{-2}$ fall off in the nuclear radiation field, we
would expect jet-induced shocks to be relatively more important on a large
scale. This is supported by the close alignments observed on this scale;
but  high-resolution kinematic studies are required to test the idea that
this gas has indeed undergone a jet-cloud interaction, by searching for
the expected kinematic signatures of such interactions. Unfortunately, the
existing spectroscopic studies  of \scite{mccarthy96} of that region, are
not of sufficient quality for this purpose.

Something similar is seen in low-redshift starburst galaxies, where
photoionization  from the central starburst source dominates in the
inner regions, and the importance of shock-ionization increases at
larger radii \cite{shopbell98}.

These results also demonstrate that the alignment effect can be
important on a large  scale, even in cases in which the alignment is
poor on a smaller scale and in which  the radio structure extends much
further than the emission-line structures.

\subsubsection{Emission-line ratios}

\begin{figure}
\centerline{ \psfig{figure=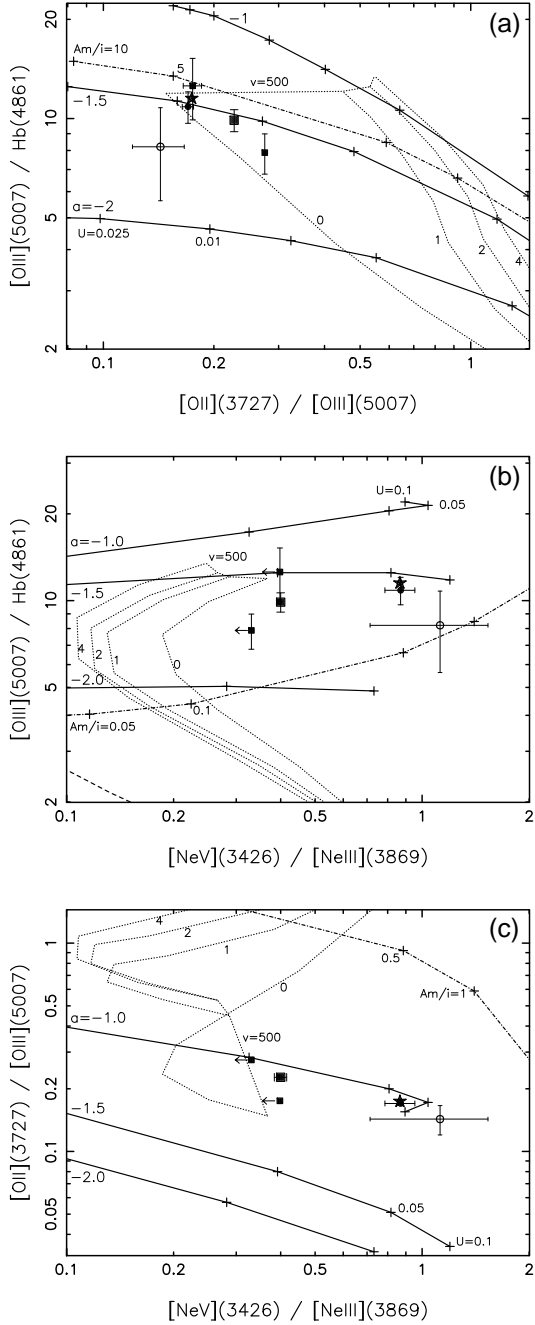,angle=-90,width=8cm}}
\caption[]{Diagnostic diagrams for different regions of
3C\hspace{0.35ex}265,  including: nuclear region (star), inner SE EELR
showing both low- (filled circle)  and high- (open circle) velocity
components, inner NW EELR (big filled square),  and outer SE and NW
EELR (small filled squares). Solid lines represent  optically-thick
power-law ($F_{\nu} \propto \nu^{\alpha}$) photoionization models
(from MAPPINGS) with $\alpha$ = --1.0, --1.5 and --2.0, and a sequence in
the ionization  parameter ($2.5 \times 10^{-3} < U < 10^{-1}$). The
dot-dash-dot line corresponds to  photoionization models including
matter-bounded clouds (10$^{-2} \leq$ A$_{M/I}  \leq$ 5) from
\scite{binette96}. The dotted lines represent shocks+precursor models
from Dopita \& Sutherland (1995, 1996)\nocite{dopita95,dopita96}, each
sequence  corresponds to a fixed magnetic parameter (B/$\sqrt{n}$ = 0,
1, 2, 4 $\mu$Gcm$^{-3/2}$)  and a changing shock velocity (150 $\leq
v_{s}\leq$~500 \kms). Note that simple shock  predictions (without a
precursor) lie outside the diagrams.}
\label{diagnostics}
\end{figure}

To study in more detail the ionization mechanism of the line-emitting
gas, we have  compared the emission-line ratios for different regions
in 3C\hspace*{0.35ex}265  with theoretical predictions of both
AGN-photoionization and shock-ionization models.  The diagnostic
diagrams are presented in Fig.~\ref{diagnostics}, and include the
following line ratios: (a) [OIII](5007)/H$\beta$
vs. [OII](3727)/[OIII](5007);  (b) [OIII](5007)/H$\beta$
vs. [NeV](3426)/[NeIII](3869); and (c)  [OII](3727)/[OIII](5007)
vs. [NeV](3426)/[NeIII](3869).

The measured line ratios were derived from the long-slit spectra along
PA\hspace*{0.5ex}145$^{\circ}$. The points correspond to the five
spatial regions  defined along the slit (see
Section~\ref{res:spec-fluxes} for details). Note that  the slit was
misaligned by $\sim$35$^{\circ}$ with respect to the radio axis.

The data are compared with various ionization models which include:
power-law  photoionization (solid lines), photoionization including
matter-bounded clouds  (dot-dash-dot line), and shocks including a
photoionized precursor (dotted lines).  Note that simple shock cooling
zone models (i.e. without a precursor region) lie  outside the
diagrams. To generate the different ionization model predictions we
followed the same procedure described in Sol{\'o}rzano-I{\~n}arrea 
et al. (2001)\nocite{carmen2001}, to which the reader is referred for
further references  and details.

By comparing the three diagnostic diagrams, we see that all of the
points fall well away from the ``pure'' shock models and within the
region occupied by the photoionization predictions, although there is
not complete consistency between the  diagrams. Note, however, that
this inconsistency does not rule out AGN-photoioinization  because the
models --- particularly the mixed-medium models --- could be tuned to
improve the fits in the diagrams. Similar problems of inconsistency,
when comparing  line ratios with ionization model predictions, have
been found for other high-redshift  radio galaxies
(Sol{\'o}rzano-I{\~n}arrea et al. 2001)\nocite{carmen2001}.  In the
case of the outer EELR (small filled squares), shock+precursor
predictions  with magnetic parameter B/$\sqrt{n}$ = 0
$\mu$Gcm$^{-3/2}$ and shock-velocity in the  range 400 $\leq v_{s}\leq$
500~\kms \ could also give a reasonable fit to those  regions; but in
all other regions the line ratios appear more consistent with
AGN-photoionization.

Note that the nucleus (star) and the low-velocity component (filled
circle) overlap  each other in the three diagrams, which indicates
that these two regions have a  similar ionization state. In contrast,
the high-velocity component (open circle)  occupies a different region
in the diagrams, with the highest ionization among all  regions,
although the errors for this component are also bigger than for the
other  points.

\subsubsection{High-velocity gas}

It seems clear that at small nuclear distances ($\lesssim$7 arcsec) the
main  ionization mechanism is photoionization by the central
quasar. However,  high-velocity gas is detected at $\sim$2.5 arcsec (20
kpc) from the centre of  3C\hspace*{0.35ex}265 (see
Section~\ref{res:kinemat}), within the ionization cone,  where the effects
of AGN illumination should dominate. The projected velocity of the
fast-moving cloud with respect to that of the surrounding gas is
approximately  1000~\kms \footnote{Note that if the high-velocity cloud is
undergoing radial  motion close to the plane of the sky, the
\emph{deprojected} radial velocity of  the cloud could be much
larger.}. Such a large velocity amplitude cannot be caused  by gravity of
a single galaxy (Tadhunter et al. 1989)\nocite{tadhunter89b}, but could be
explained in terms of acceleration by the interaction with the radio
structures. \scite{tadhunter91}  suggests that such large velocity shifts
are due to cocoons of material expanding  around the radio jets, perhaps
remnants of the bow-shocks associated with the  passage of the jet. This
is strongly supported by the location of the high-velocity  cloud, which
lies close to the radio axis (see Fig.~\ref{lhvel}). Although acceleration
by AGN- or starburst-induced winds could be considered as an alternative
mechanism  (e.g. \pcite{heckman90}), such winds would be expected to
produce a broad outflow right  across the cone, and not localised on the
radio axis.

It is surprising that a galaxy such as 3C\hspace*{0.35ex}265, in which
the radio source is on a much larger scale that the emission-line
structures, presents signs of jet-cloud interactions. However,
3C\hspace*{0.35ex}265 is not the only large radio source in which these
interactions have been detected. Kinematic evidence for shock-acceleration
has been also found in two other high-redshift radio galaxies
(3C\hspace*{0.35ex}34 and 3C\hspace*{0.35ex}330), in which the radio hot
spots have passed well beyond the EELR (Sol{\'o}rzano-I{\~n}arrea et al. 2001)\nocite{carmen2001}.

Although the high-velocity cloud may have been shock-accelerated, that
does not necessarily mean that the cloud is shock-ionized. It could be
accelerated by jet-induced shocks but still photoionized by the central
active  nucleus. In either case, if we believe the high-velocity cloud to
be accelerated  as a result of the interaction with the radio structures,
the density of the cloud will increase because of the compression effect
of the shocks (a factor of 10 to  1000;
e.g. \pcite{begelman89,komissarov94}), and therefore a lower ionization
parameter  than that of the surroundings is expected. However, as
discussed in the previous  section, this is not observed. The
high-velocity gas appears to have similar, if not higher, ionization state
than the emission-line regions in the vicinity.

If the high-velocity gas is actually the result of the interaction
with the radio  components, then how can we explain the
high-ionization state and narrow linewidths  observed in the
line-emitting gas? Assuming that the regions behind the shock front
are destroyed, we would expect to measure weaker low-ionization
emission lines,  which are emitted mainly by the cooling post-shock
clouds, resulting in a  higher-than-expected observed ionization
state. Thus, `matter-bounded shocks',  previously considered for
high-redshift radio galaxies  (Sol{\'o}rzano-I{\~n}arrea et
al. 2001)\nocite{carmen2001}, could be another  possible explanation
for the high-velocity cloud in 3C\hspace*{0.35ex}265.

Alternatively, since the cloud lies close to the direction of the
radio axis, it  could fall within the narrow cone
($\lesssim$10$^{\circ}$ half-opening angle)  defined by the blazar
beam of the relativistic jets. The higher ionizing flux density of the
beam --- about an order of magnitude greater than that of the unbeamed
component \cite{browne87} --- might dominate over the compression
effect of  the shocks, therefore increasing the ionization state of
the high-velocity cloud.  However, given that the compression across a
strong radiative shock could be a  factor of $>$100
(e.g. \pcite{begelman89,komissarov94}), it is not clear that the boost
in ionizing flux density due to the blazar beam would be enough to
produce the high ionization state in the high-velocity gas.

It is notable that kinematic disturbed components with high ionization
state have been detected in the nuclear regions of several other
objects, including the extended radio source Cygnus\hspace*{0.35ex}A
\cite{tadhunter91}, and the compact radio sources
PKS\hspace*{0.35ex}1549-79 \cite{tadhunter2001},
PKS\hspace*{0.35ex}1345+12 \cite{grandi77} and 3C\hspace*{0.35ex}48
\cite{chatzichristou99}.  These three latter sources present
high-ionization emission lines that are unusually broad and shifted
relative to the low-ionization emission lines.  In the case of
Cygnus\hspace*{0.35ex}A, a high-velocity component, with similar or
higher ionization state than the unshifted component, is detected at
$\sim$1 arcsec (1.5 kpc) NW of the radio core and located close to the
radio axis, just like 3C\hspace*{0.35ex}265
\cite{tadhunter91,tadhunter94,stockton94}.

In contrast with these sources, detailed studies of the jet-cloud
interactions on a larger scale (20 -- 50 kpc) in the powerful radio
galaxy PKS\hspace*{0.35ex}2250--41  have revealed multicomponent
emission-line kinematics: a low-ionization broad  component,
representing material cooling behind the shock front; and a
high-ionization  narrow component, which represents the AGN- or
shock-photoionized precursor gas  \cite{villarmartin99b}. Our
long-slit spectra of 3C\hspace*{0.35ex}265 reveal  relatively narrow
linewidths ($<$250~\kms) for the high-velocity emission-line gas,
similar to those of the low-velocity component (see
Section~\ref{res:spec-kin}). However, in this case, unlike for
PKS\hspace*{0.35ex}2250--41, the narrow high-velocity  component
cannot represent the photoionized precursor gas, given the high
velocity of  $\sim$1000 \kms \ at which the cloud is moving relative
to the surrounding gas.

Overall, the  existence of fast-moving  gas along the radio  axis with
higher  ionization  states  than  the  surrounding  emission-line  gas
remains a mystery. No explanation  for the high ionization state seems
entirely satisfactory, and further  investigation is required for this
issue.

\subsection{Intrinsic gas structure}

3C\hspace*{0.35ex}265  is   an  extreme  radio  galaxy   in  terms  of
luminosity,  morphology  and  emission-line  properties.  Its  bizarre
optical  morphology,  completely surrounded  by  emission regions  and
companion  galaxies,  revealed by HST images, suggests  that
3C\hspace*{0.35ex}265  is  undergoing the  process  of merging
\cite{longair95}.  More recent  analysis of  the environment around 3C265
show that there are indeed companion galaxies surrounding 3C265,  and
reveal  that  the  galaxy lies  in  a cluster  environment
\cite{pbest2000}.

The continuum image, presented in Fig.~\ref{o2o3cont} (c), shows that the
closest companion galaxy, to the north-north-west of
3C\hspace*{0.35ex}265, aligns along the axis of the highest surface
brightness [OII], [OIII] and UV-continuum emission elongation
(PA$\sim$150$^{\circ}$).  In Section~\ref{res:ioniz} it was shown that the
ionization cone of the AGN in 3C\hspace*{0.35ex}265 is indeed  aligned
with the radio axis, therefore the optical/UV emission  elongation must be
intrinsic, since there is no reason why the gas  would not be illuminated
symmetrically. This suggests that the gas  and dust present along  the
position  angle   of  the  optical/UV  elongation  may  be associated with
an interaction/merger involving 3C\hspace*{0.35ex}265 and this companion.
Moreover, both the increase of the covering factor due to the extra amount of
gas along this direction, and the fact that 3C\hspace*{0.35ex}265 is among the
most powerful radio sources known (e.g. \pcite{laing83}) may explain why 
its emission-line luminosity is so extreme.

It  is  also intriguing  that  the next  closest  galaxy  in terms  of
projected distance from the nucleus --- 11 arcsec (90 kpc) to the east
of  the  nucleus  --- is  aligned  with  the  PA  of the  large  scale
($\gtrsim$9  arcsec;  75  kpc)  gas  on  both  sides  of  the  nucleus
(PA$\sim$120$^{\circ}$).  If  more  detailed observations  demonstrate
that this is a genuine companion galaxy, then it is plausible that the
larger  scale emission-line  structure  also represents  a gas  stream
associated with  a merging  companion. In this  case, the  large scale
elongation  in  the gas  distribution  along  the  radio axis  may  be
intrinsic, rather than a consequence of jet-cloud interactions.

Similarly, images of other powerful radio galaxies reveal companion
galaxies aligned  along the direction of the aligned optical/UV
emission. Examples include: 3C\hspace*{0.35ex}65,
3C\hspace*{0.35ex}266,  3C\hspace*{0.35ex}324, 3C\hspace*{0.35ex}356
and 3C\hspace*{0.35ex}368 (\pcite{dunlop93};  Best et al. 1996,
1997).\nocite{best96,best97}

This suggests that the highest surface brightness structures in radio
galaxies may  be aligned along a preferred direction, defined by
mergers with companion galaxies.  This is in agreement with the idea
proposed by \scite{west94}, who suggests that  powerful radio galaxies
at high redshifts are formed by a hierarchical, highly-anisotropic
merging processes.
 
\section{Conclusions}

We report the results from a study of the morphology, kinematics and ionization
structure of the powerful radio galaxy 3C\hspace*{0.35ex}265.

Our results show that different ionization mechanisms may dominate on
different  scales. We find that close to the nucleus the dominant
ionization mechanism is  AGN-photoionization. This is supported by the
detection of the broad ionization  cones predicted by the unified
schemes. In contrast, as we move to large distances  from the nucleus,
jet-cloud interactions become the main ionization mechanism of the
line-emitting gas, and the gas appears more closely aligned with the
radio axis. The  presence of a high-velocity cloud, located close to
the passage of the radio jet and  centred at approximately 2.5 arcsec
(20 kpc) from the nucleus, shows that the effects  of jet-induced
shocks are still important in the nuclear regions. However, it is
particularly puzzling that this fast-moving gas presents a high
ionization state; if  the cloud is believed to have been
shock-accelerated, it would be expected to have  lower ionization
state than the surroundings, but this is not observed. Clearly,  more
data are required to investigate this issue.

In addition, our results suggest that the intrinsic distribution of
the gas close to the nucleus ($\lesssim$60 kpc) in
3C\hspace*{0.35ex}265 is aligned along a preferred direction
(misaligned with the  radio axis), which may be the result of a
highly-anisotropic merging  process. On the other hand, at larges
distances from the nucleus, where jet-cloud interactions dominate, the
low-ionization emission-line regions align along the  direction of the
radio axis.

Overall, our results demonstrate that a combination of mechanisms is
required in  order to explain the observed properties of the
line-emitting gas in the haloes of  radio galaxies. This confirms our
previous results presented in  Sol{\'o}rzano-I{\~n}arrea et
al. (2001), which show evidence for shock-acceleration  in the EELR of
all the galaxies in a small sample, even those in which the shock
fronts have passed the EELR and AGN-illumination would have been
expected to  dominate.

\section*{Acknowledgments}

This work is based on observations made at the Observatorio del Roque de
los  Muchachos, La Palma, Spain. The long-slit spectra used for this paper
were taken  from the ING Archive. We thank Philip Best for reading the
manuscript and giving  useful suggestions. CSI acknowledges a White Rose
studentship. This research has made use of the NASA/IPAC Extragalactic
Database (NED) which is operated by the Jet Propulsion Laboratory,
California Institute of Technology, under contract with the National
Aeronautics and Space Administration. We thank the anonymous referee for
helpful comments.

\bibliographystyle{mnras}
\bibliography{reference}

\end{document}